\documentclass[aps, prl, twocolumn, showpacs, superscriptaddress, nofootinbib]{revtex4}
\usepackage{amsmath, amssymb, graphicx, natbib, pdfpages}

\begin{document}

\title{Target search on a dynamic DNA molecule}

\author{Thomas Sch\"otz}
\affiliation{Arnold Sommerfeld Center for Theoretical Physics and Center for Nano Science, University of Munich, Theresienstr. 37, D-80333 M{\"u}nchen, Germany}
\author{Richard A. Neher}
\affiliation{Max-Planck-Institute for Developmental Biology, Spemannstr. 35, 72076 T\"ubingen, Germany}
\author{Ulrich Gerland}
\email{gerland@lmu.de}
\affiliation{Arnold Sommerfeld Center for Theoretical Physics and Center for Nano Science, University of Munich, Theresienstr. 37, D-80333 M{\"u}nchen, Germany}

\date{\today}

\begin{abstract}
We study a protein-DNA target search model with explicit DNA dynamics applicable to {\it in vitro} experiments. We show that the DNA dynamics plays a crucial role for the effectiveness of protein ``jumps'' between sites distant along the DNA contour but close in 3D space. A strongly binding protein that searches by 1D sliding and jumping alone, explores the search space less redundantly when the DNA dynamics is fast on the timescale of protein jumps than in the opposite ``frozen DNA'' limit. We characterize the crossover between these limits using simulations and scaling theory. We also rationalize the slow exploration in the frozen limit as a subtle interplay between long jumps and long trapping times of the protein in ``islands'' within random DNA configurations in solution.
\end{abstract}


\maketitle

The quantitative characteristics of proteins searching for their specific target sites on long DNA molecules has become a paradigmatic question of biological physics \cite{Halford_NAR_04, Riggs_JMolBiol_70, Richter_BiophysChem_74, BWvH_81}. The question is of considerable biological interest, since search processes of this type are key steps in cellular functions. For instance, in signal transduction, a protein belonging to the large class of transcription factors conveys an external signal and triggers the appropriate genetic response by binding to specific target sites on the genomic DNA. Similarly, restriction enzymes, used by bacteria to fight invading viruses, search for cleavage sites marked by specific DNA sequences. It is generally assumed that the target search mechanism has been optimized by evolution, due to selective pressure for fast signaling and rapid responses in competitive environments. 
From the physics perspective, the protein-DNA target search is a complex but tractable stochastic process that combines basic aspects of Brownian motion, polymer physics, and information theory \cite{Stormo_TIBS_98, Gerland_PNAS_02, Bruinsma_PhysicaA_02, Brockmann_PRL_03, Slutsky_BiophysJ_04, Coppey_BiophysJ_04, Lomholt_PRL_05, Hu_BiophysJ_06, Hu_PRE_07, Wunderlich_NAR_08, Sheinman_PhysBiol_09}. 
Experimentally, the search process can be probed on the single-molecule level {\it in vitro} \cite{vandenBroek_PNAS_08}, and even {\it in vivo} \cite{Elf_Science_07}. 

Early {\it in vitro} experiments \cite{Riggs_JMolBiol_70} indicated that the association rate of {\it lac} repressor to its target site embedded in short pieces of DNA is faster than the diffusion limit, $k_{a}=4\pi D b$, for a direct binding reaction with diffusion constant $D$ and reaction radius $b$. 
Inspired by Adam and Delbr\"uck's idea that reduction of dimensionality is a generic way to enhance reaction rates \cite{Delbrueck_68}, Richter and Eigen \cite{Richter_BiophysChem_74} interpreted these experiments with a two-step mechanism where 3D diffusion and non-specific association to DNA is followed by 1D diffusive sliding into the target site. In a seminal series of papers \cite{BWvH_81}, Berg, Winter, and von~Hippel then established much of what is known today about the protein-DNA search kinetics. They experimentally varied the non-specific binding strength via the ion concentration, identified an optimum where the search is fastest, and explained the behavior in a theoretical analysis. 

\begin{figure}[b]
\centering
\includegraphics[width=7.5cm]{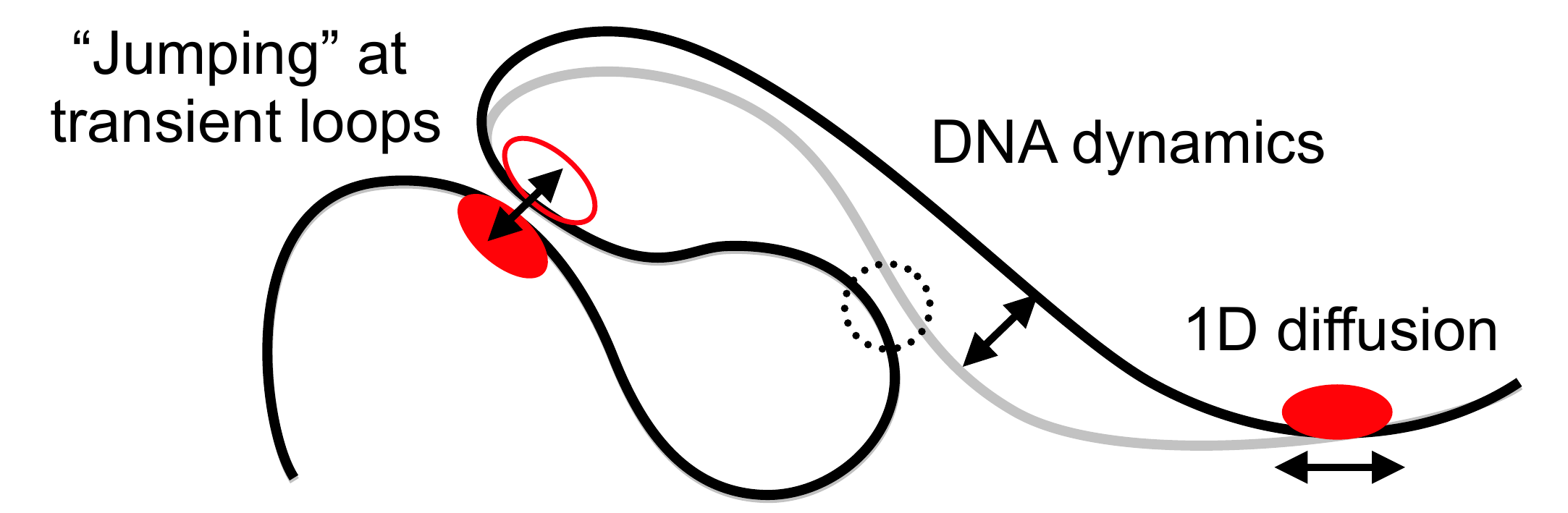}
\caption{Illustration of the target search by sliding (1D diffusion) and jumping on a dynamic polymer.} 
\label{fig:model}
\end{figure}

The existence of an optimum reflects a generic tradeoff in search processes for hidden targets \cite{Benichou_PRL_05}: A stochastic local search is exhaustive but redundant; interrupting the search by phases of rapid movement to new territory is a time investment that pays off by reducing the redundancy. The optimal fraction of time spent in each of the two ``modes'' depends on the statistical characteristics of the search mechanism. The simplest scenario, where proteins slide diffusively along the DNA, dissociate spontaneously, and randomly reattach at uncorrelated positions, leads to an optimum where, on average, only half of the proteins are bound somewhere on the DNA and the other half is in solution \cite{BWvH_81}. Physically, this is best understood \cite{Slutsky_BiophysJ_04} in terms of the typical dwell times of a protein in the sliding mode, $\tau_{s}$, and in the dissociated state, $\tau_{d}$. The latter should be regarded as a fixed parameter, set by cell size and composition, whereas $\tau_{s}$ can be adapted by molecular evolution of the DNA-binding domain of the protein (to adjust the non-specific affinity). If $\tau_{s}<\tau_{d}$, the protein spends too little time searching, while if $\tau_{s}>\tau_{d}$, the search is too redundant; the search is fastest when they are equal. 

However, in bacterial cells, well studied transcription factors are bound to DNA $\gtrsim\!90\%$ of the time \cite{Stormo_TIBS_98}. This fact has drawn attention to the `intersegment transfer' \cite{BWvH_81, Lomholt_PRL_05, Hu_PRE_07, Sheinman_PhysBiol_09} of proteins within the same DNA molecule, between sites close in space but distant along the contour. Potentially, this process can destroy the redundancy of the 1D search without the price of interrupting it by long excursions into the solvent. The term was introduced for proteins with two DNA-binding domains and refers to a process during which the protein never detaches from the DNA; a similar transfer but with a brief unbound period is referred to as `hopping' \cite{BWvH_81}. In both cases, the essential difference to the uncorrelated random reattachment discussed above is the correlated nature of the process: Transfer does not occur with equal probability to every site on the DNA, but to ``linked'' sites. Here, we simply refer to both processes as `jumping'. 

The interplay of protein sliding and jumping leads to intricate search dynamics. An analytical study \cite{Lomholt_PRL_05} considered the effect of jumps using the fractional Fokker-Planck equation \cite{Metzler_PhysRep_00}, which assumes that consecutive jumps are uncorrelated, i.e. that the DNA configuration randomizes between two jumps. In contrast, a numerical study of sliding and jumping on a random but frozen contour \cite{Sokolov_PRL_97} showed that correlations between jumps drastically alter the dynamics, leading to ``paradoxical'' quasi-diffusive behavior instead of super-diffusion along the contour. Specifically, the distribution of the protein on the DNA exhibits characteristic heavy tails even though its width increases only diffusively. These findings, and the fact that the dynamics of real DNA is neither frozen nor annealed over the relevant range of $\mu s$ to $s$ timescales \cite{BWvH_81}, call for an analysis of target search on a dynamic DNA, see Fig.~\ref{fig:model}. Here, we characterize the crossover between the frozen and the annealed regime using simulations and scaling theory. We then study the mechanism whereby correlated jumps create the paradoxical behavior in the frozen limit.

{\it Model.---}
To make the problem tractable, we describe the DNA contour as a path of $L$ segments on a simple cubic lattice, and generate its conformational dynamics with a kinetic Monte Carlo scheme based on a generalized Verdier-Stockmayer move set \cite{Binder_2000} with moves for kinks, chain ends, and crankshafts, see Fig.~S1. These moves, carried out at rate $k_\mathrm{D}$, implement Rouse dynamics on a lattice for an ideal chain (no self-avoidance). We describe a protein as a point particle on the lattice, which diffuses along the DNA contour at rate $k_\mathrm{p}$. If another DNA segment passes through the same point, the protein can randomly jump to it (at the same rate $k_\mathrm{p}$, for simplicity). We focus on the limit of strong DNA binding without explicit 3D diffusion of the protein (although jumps may involve 3D diffusion, as discussed above). As initial condition, we use a random DNA configuration with the protein on the central segment. Clearly, the configuration of the DNA inside a bacterial cell is not random, due to genome packaging and confinement, but a random configuration is an interesting starting point for exploration of the physical principles, and mimics the situation of {\it in vitro} experiments. 

\begin{figure}[tbp]
\centering
\includegraphics[width=8cm]{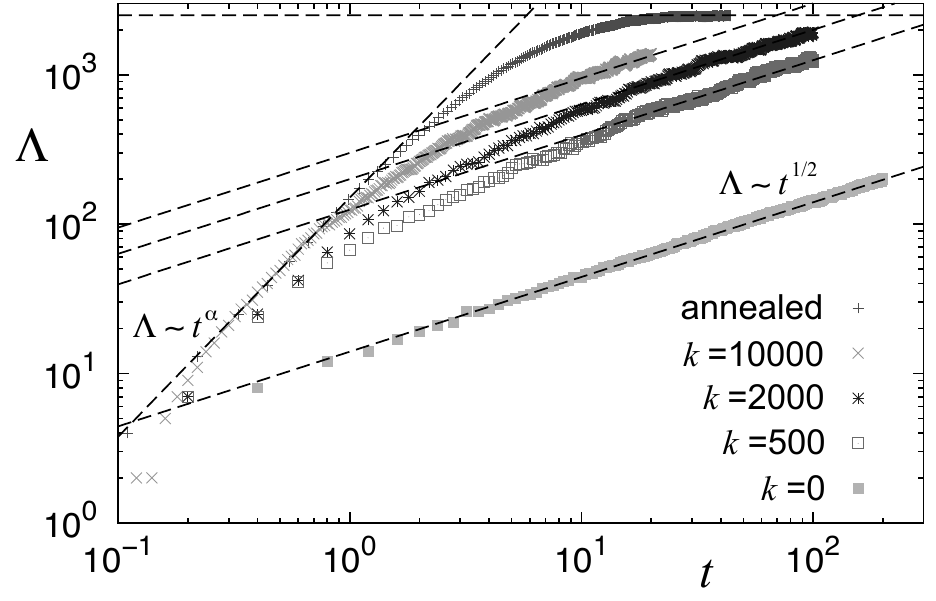}
\caption{\label{fig:crossover} 
Time evolution of the width $\Lambda$ of the protein distribution $P(s,t)$ for different kinetic ratios $k=k_\textrm{D}/k_\textrm{p}$. A crossover from super-diffusive to quasi-diffusive dynamics occurs for finite $k$. }
\end{figure}

{\it Transport.---}
To characterize how a protein explores the search space, we study the time evolution of its probability distribution $P(s,t)$ along the DNA contour ($0\le s\le L$). Fig.~\ref{fig:crossover} plots its width $\Lambda(t)$, defined as the interquartile range $\Lambda=I^{-1}(\frac{3}{4})-I^{-1}(\frac{1}{4})$ of the cumulative distribution $I(y)=\int_0^y {\mathrm d}s\,P(s,t)$, for different kinetic ratios $k=k_\textrm{D}/k_\textrm{p}$. We obtain $P(s,t)$ by averaging over $\ge\!10^{3}$ simulations, with $L=5000$ and different initial DNA configurations. In the `quenched limit' $k\to 0$ (squares), the protein moves on a frozen contour, and the width grows quasi-diffusively with time, $\Lambda\sim t^{1/2}$, despite the long-range jumps along the contour and a heavy tail of the distribution $P(s,t)$ at fixed $t$ \cite{Sokolov_PRL_97}. In the opposite `annealed limit' $k\to\infty$ (crosses, obtained by randomly drawing a new DNA configuration after each jump), the distribution initially spreads super-diffusively along the contour, $\Lambda\sim t^{\alpha}$ (here: $\alpha\approx 1.7$). The width saturates at $\Lambda\to L/2$ as the protein explores the entire DNA. In the regime of intermediate $k$, which is relevant in most experimental situations, $\Lambda(t)$ displays a crossover from super- to quasi-diffusive scaling. The curves for different $k$ show that the crossover timescale $\tau_{c}$ increases with $k$. 

For large $k$, the connectivity of the DNA meshwork on which the protein moves changes rapidly, such that successive jumps are uncorrelated (they occur on different link sets). One can then describe the dynamics by the average jump probability $P(s,s')$ from site $s$ to site $s'$, which is physically determined by the DNA looping probability. For an ideal chain, this probability decays as $|s-s'|^{-3/2}$ for large loops, before it is cut off by the finite DNA length. When successive jump lengths are independently drawn from this distribution, the typical distance $\Lambda$ from the initial position is dominated by the largest jump, which grows with the number of jumps ($\sim t$) as $\Lambda(t)\sim t^{2}$ \cite{Bouchaud_PhysRep_1990}. Indeed, our numerical exponent $\alpha$ approaches 2 at large $L$ (data not shown). However, what does the transport $\Lambda(t)$ imply for the target search process?

{\it Search time.---} 
Without a guiding ``funnel'', no search process can be faster than linear exploration. A faster than linear $\Lambda(t)$ leads to ``sloppy search'' \cite{Lomholt_PRL_05} where patches dispersed over the entire contour are explored before the target is located. This is precisely what is required to break the redundancy of 1D diffusion, suggesting that jumping is an effective mechanism that could replace 3D diffusion in the annealed regime. On the other hand, we expect that jumping is ineffective in the frozen limit, as it leads only to quasi-diffusive spreading along the DNA. To study the target search on a dynamic DNA explicitly, we performed simulations with a target site placed at different distances from the initial protein position. Fig.~S2A shows that the search indeed takes increasingly longer as the DNA dynamics is slowed. Fig.~S2B shows that the strong dependence of the search time on the initial distance to the target (at $k=0$) becomes weaker as $k$ is increased, see caption for details. 

It will require single-molecule experiments of the type of \cite{Elf_Science_07} (but under controlled {\it in vitro} conditions) to find out which regime of $k$ values is biologically most relevant. However, a rough estimate, based on the experimental relaxation time of $\tau=30$~s for the contour of a $L=43\,\mu$m DNA fragment and the experimental scaling law $\tau\sim L^{1.65}$ \cite{Perkins_Science_94}, indicates that on the ms-timescale of protein jumps \cite{BWvH_81}, only short DNA segments will be equilibrated. We therefore expect that neither the annealed nor the frozen limit, but the crossover regime will be most relevant experimentally.

{\it Scaling of the crossover.---}
To understand the physics of the crossover regime within our model, we apply a scaling argument to the interplay of DNA and protein dynamics: A DNA segment of length $\ell$ equilibrates on a time scale $\tau\sim\ell^{2}$ (Rouse dynamics). Within a time $\tau$ after a protein docks onto the DNA and starts exploring, it typically visits a DNA stretch $\Lambda(\tau)$. During this time, a DNA segment of size $\ell\sim(k_\mathrm{D}\tau)^{1/2}$ equilibrates. Superdiffusive protein transport results as long as $\Lambda(\tau)<\ell$, however the fast growing $\Lambda(t)\sim (k_\mathrm{p}t)^\alpha$ quickly outruns the ``equilibration blob'', and the passing point marks the crossover to the quasi-diffusive regime. With $\alpha=2$, this crossover timescale $t_c$ then depends on the kinetic ratio $k$ as $k_\mathrm{p}t_c\sim k^{1/3}$. Our simulations cannot explore a wide range of $k$ values due to computational cost and do not allow a precise determination of this scaling (however, the scaling exponent that best describes our limited data deviates only by 0.08 from the expected value $1/3$, see Fig.~S3). The small numerical value of the exponent leads to a broad crossover as a function of $k$, again suggesting that neither the annealed nor the frozen limit is experimentally attainable.

\begin{figure}[t]
\centering
\includegraphics[width=8.6cm]{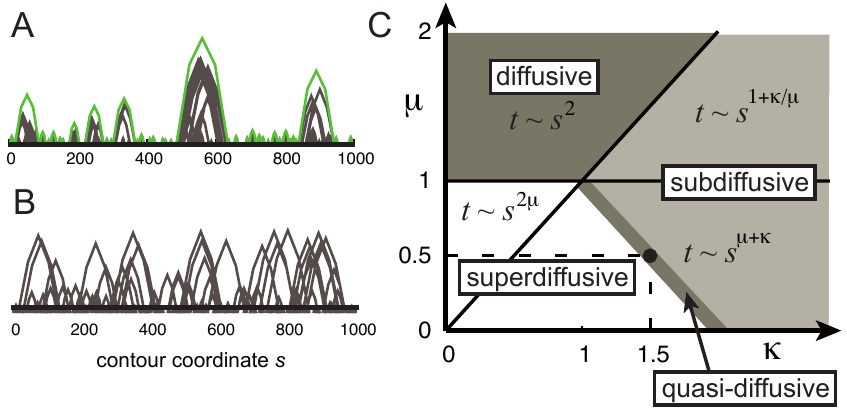}
\caption{The link diagram for a typical DNA conformation (A) is separable into islands (green). Random reshuffling of the same links destroys the islands (B). A toy model for transport on the island structure leads to the dynamical phase diagram (C), which explains the quasi-diffusive regime as a cancellation of the effect of traps and long-range jumps. 
}
\label{fig3}
\end{figure}

{\it Quenched limit.---}
To obtain a better understanding of the mechanism responsible for the slow down of the search, we focus on the quenched limit. When first reported \cite{Sokolov_PRL_97}, the quasi-diffusive transport was attributed to correlation effects. However, what is the nature of these correlations and how do they render the long-range jumping process quasi-diffusive? We distinguish two types of correlations, which we refer to as {\em temporal} and {\em spatial}. On a static DNA, a protein can use the same links multiple times, leading to temporal correlations. Additionally, the positions of different links are spatially correlated, since an existing link strongly enhances the probability to find another link nearby (e.g. a loop in the DNA favors further contacts within the loop). 
To separate the effect of temporal and spatial correlations, we destroy the latter by choosing a new random starting point for each link while conserving its arc length $|s-s'|$. The protein transport on such reshuffled link sets is super-diffusive as revealed by simulations shown in Fig.~S4. Hence temporal correlations alone are not sufficient to cause the quasi-diffusive behavior. A simple argument makes this plausible: If the region visited by the protein grows super-diffusively as $\Lambda(t)\sim t^2$, the protein visits only a fraction $\sim1/t$ of the sites within $\Lambda$. Since it sees each site $\mathcal{O}(1)$ times, it mostly uses novel links and the persistence of links is unimportant.

{\it Islands.---}
A striking consequence of the spatial correlations is revealed in Fig.~\ref{fig3}A, where all links in a typical DNA configuration are depicted as arcs. The arcs cluster into ``islands'' with many internal links but no links between islands. These islands disappear when the same links are randomly placed on the DNA, see Fig.~\ref{fig3}B. Intuitively, it is clear that the existence of islands slows the exploration of the DNA, since the protein can move from one island to another only by sliding. In fact, if the islands had a well-defined typical size $\overline{s}$, the protein dynamics would be diffusive on long scales $s\gg\overline{s}$. However, the problem is more intricate, since the distribution of island sizes has the same heavy tail $p(s)\sim s^{-3/2}$ as the link length distribution, see Fig.~S5. Nevertheless, the existence of islands is a crucial clue; we show below that it leads to a dynamics that can be described by a 1D transport model with traps and long-distance jumps. To this end, we first note two essential transport properties of islands: 
(i) Due to the internal links, the position of a protein is rapidly randomized within an island, such that for most starting positions within an island, it leaves the island with nearly the same probability to each side, see Fig.~S6. 
(ii) The typical trapping time within an island scales as $\tau\sim s^{3/2}$ with the island size, see Fig.~S7.  

Given these properties, we consider protein transport on an array of islands with sizes $s_i$ drawn from the distribution $p(s)$. Each island has an associated trapping time $\tau_{i}(s_{i})$. It will be instructive to allow for adjustable exponents $\mu$ and $\kappa$ in the scaling behavior, $p(s)\sim s^{-1-\mu}$ and $\tau \sim s^{\kappa}$. Combining these relations, we obtain a distribution of trapping times $w(\tau)\sim\tau^{-1-\mu/\kappa}$, since $w(\tau){\mathrm d}\tau=p(s){\mathrm d}s$. The transport behavior of the protein in island space is then determined by the ratio of the exponents: Using the first passage time calculus \cite{Gardiner_2004}, the typical time needed to move over $n$ islands is 
\begin{equation}
\label{eq:mfpt}
T \sim n \sum_{i=1}^n \tau_i\sim 
\left\{\begin{array}{ll}
n^{1+\frac{\kappa}{\mu}}&\textrm{ for }\qquad \kappa>\mu \\
n^2 & \textrm{ for }\qquad \kappa<\mu
\end{array}
\right. \;,
\end{equation}
with the sum dominated by the largest term for the case $\kappa>\mu$ while a typical trapping time exists for $\kappa<\mu$. To map the dynamics in island space back onto the DNA, note that the total DNA length $S$ of $n$ islands scales as 
\begin{equation}
\label{eq:Sofn}
S(n) \sim
\left\{
\begin{array}{ll}
n^{1/\mu}&\textrm{ for }\qquad \mu<1\\
n & \textrm{ for }\qquad \mu>1
\end{array}
\right. \;,
\end{equation}
as $S$ is dominated by the largest island for $\mu<1$. Combining (\ref{eq:mfpt}) and (\ref{eq:Sofn}) yields the transport behavior along the DNA, i.e. the typical time to travel a given distance. Fig.~\ref{fig3}C shows the phase diagram spanned by the exponents $\mu$ and $\kappa$. It exhibits four different regimes. For $\mu>1$, the distribution of island sizes has a well defined mean and no super-diffusion can occur, but sub-diffusive dynamics results when the trapping time distribution has a sufficiently heavy tail ($\mu<\kappa$). If $\mu<1$, the dynamics is super-diffusive unless long trapping times in islands compensate for long jumps. In particular, $t\sim s^{\mu+\kappa}$ for $\mu<\kappa$, which includes the case of interest here, where the two exponents precisely add up to 2, rationalizing quasi-diffusion in the quenched limit. Within our more general island model, a whole line of points exists where the dynamics is quasi-diffusive. In contrast, for the protein transport on the DNA contour, $\mu$ and $\kappa$ are not independent, since they are both related to the statistics of the network topologies created by the DNA conformations. Why this leads to $\mu+\kappa=2$ remains to be understood.

{\it Conclusion.---}
We analyzed the transport and search of proteins on a dynamic DNA contour. We showed that the highly correlated nature of the protein dynamics persists over a broad range of our dimensionless dynamic parameter $k=k_\textrm{D}/k_\textrm{p}$ and significantly slows down the search process. Our findings imply that under the {\it in vitro} conditions of our model, protein jumping is effective as a mechanism to destroy the redundancy of a diffusive 1D search only if the DNA dynamics is sufficiently fast compared to the timescale between protein jumps or if many proteins search in parallel. Of course, the {\it in vivo} situation is complicated by many additional factors, such as the non-random conformation and the confinement of the DNA. We also found that the ``paradoxical'' quasi-diffusive dynamics in the quenched limit \cite{Sokolov_PRL_97} can be viewed as a subtle cancellation of the effect of traps and long-distance jumps. The interplay between traps, jumps, and memory in 1D transport is an intricate problem in statistical mechanics \cite{stat_mech}. The protein-DNA system naturally displays a nontrivial interplay and surprisingly is tuned to a critical point in our dynamical phase diagram.

\paragraph*{Acknowledgements. ---}
We acknowledge useful discussions with Yariv Kafri, Joseph Klafter, Ralf Metzler, Igor Sokolov and funding by the DFG via NIM.

\clearpage


\section*{Supplementary material}

# Supplementary material to: Target search on a dynamic DNA molecule

Thomas Schötz,[1] Richard A. Neher,[2] and Ulrich Gerland[1]

[1] *Arnold Sommerfeld Center for Theoretical Physics and Center for Nano Science,*
*University of Munich, Theresienstr. 37, D-80333 München, Germany*

[2] *Max-Planck-Institute for Developmental Biology,*
*Spemannstr. 35, 72076 Tübingen, Germany*

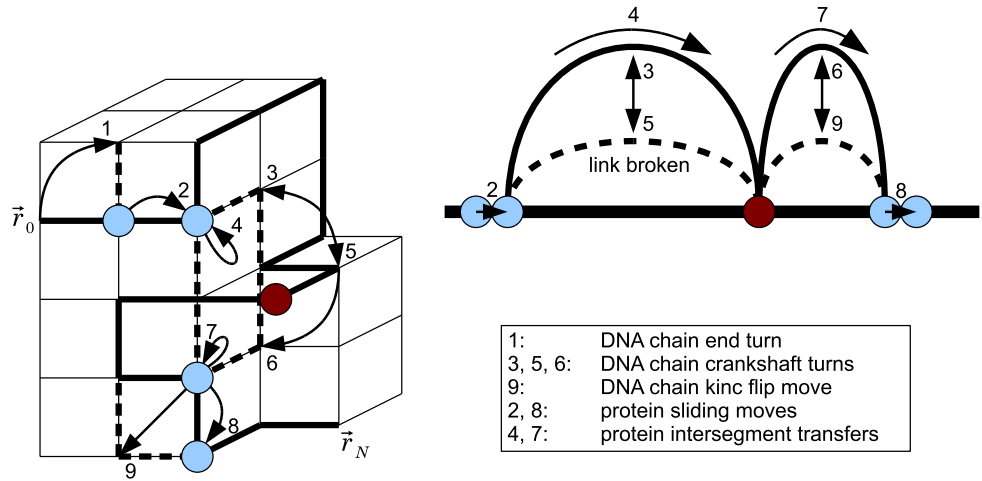

FIG. S1: Illustration of the move set of our kinetic Monte Carlo scheme. The DNA chain is represented by a path on a cubic lattice (left). The protein is represented by a point particle which moves at rate $k_\mathrm{p}$, either by randomly sliding along the chain contour or by jumping to another segment of the chain at the same position. The link diagram representation (right) has the DNA contour stretched out to a line and indicates possible jumps by arcs. Links can be created or destroyed by the Rouse dynamics of the DNA, which is implemented with a generalized Verdier-Stockmayer move set allowing for kink flips, turns at the chain end, and crankshaft moves. Each move is carried out at the rate $k_\mathrm{D}$.

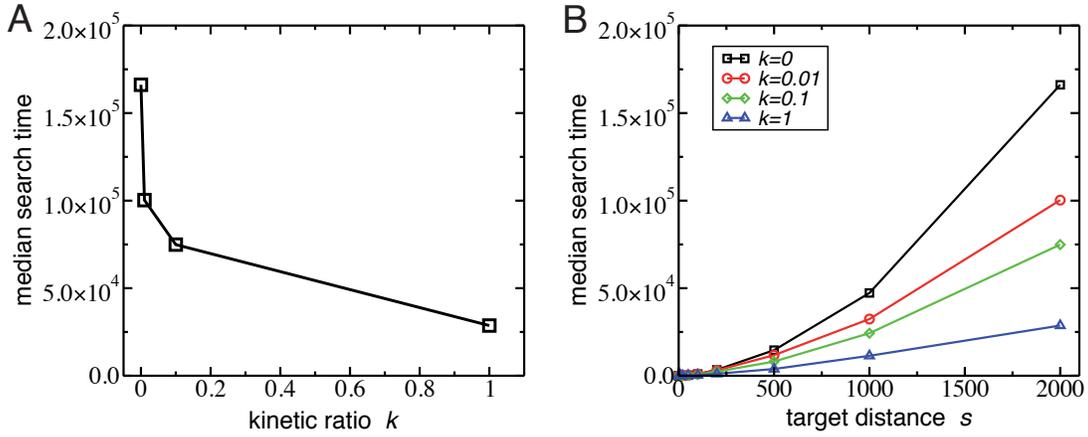

FIG. S2: Simulation results for the target search on the dynamic DNA chain (of length $L = 5000$). The protein was initially placed in the center of the chain and the time was measured until it first arrived at a target site placed a distance $s$ away. These simulations were performed 1000 times, using different initial polymer configurations, to determine the median of the search time, which represents a typical search time. Both $s$ and the kinetic ratio $k = k_D/k_p$ of DNA to protein moves were varied. The simulation of the target search on the dynamic chain is computationally expensive, which limits our range of $k$ values. (A) The typical search time to a target site at distance $s = 2000$ as a function of the kinetic ratio $k$. A substantial increase of the search time with slowing DNA kinetics is apparent. (B) The typical search time plotted against the distance of the target site, for the different kinetic ratios. The dependence of the search time on the distance $s$ becomes weaker as $k$ is increased (however, a significant dependence on $s$ remains for all our $k$ values).

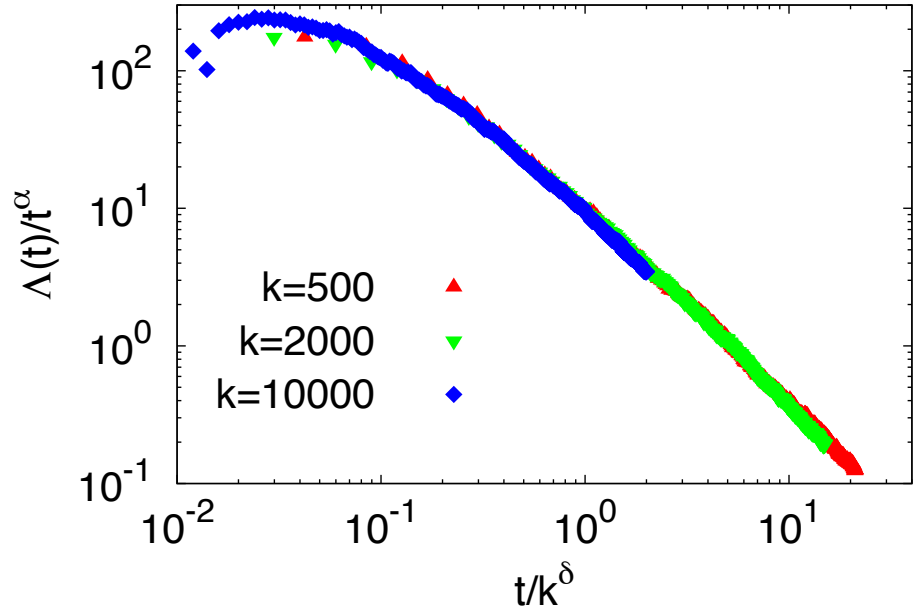

FIG. S3: Data collapse to extract an estimate of the scaling behavior of the crossover between super-diffusive and quasi-diffusive dynamics. The three curves from Fig. 2 (main text) with different finite kinetic ratios $k$ can be collapsed onto each other by rescaling of the axes (here we have used the asymptotic value of 2 for the exponent $\alpha$). The best collapse is obtained when the time is rescaled as $t/k^\delta$ with $\delta$ around 0.25. This exponent deviates from the theoretically expected value of 1/3, however the deviation is not significant given the finite size effect of our simulations.

4

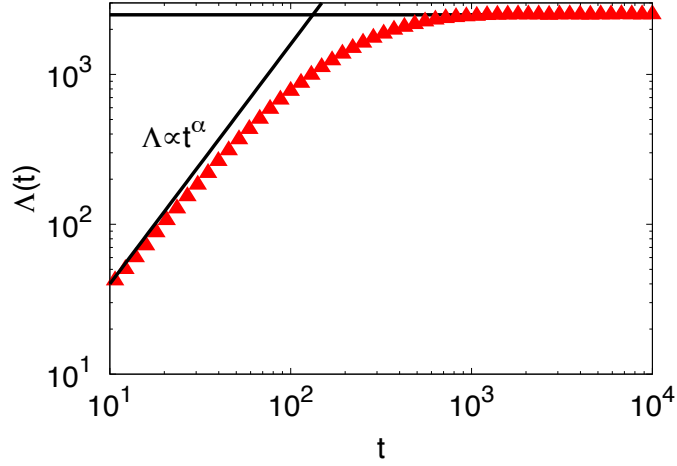

FIG. S4: Kinetic Monte Carlo simulation of the protein dynamics on reshuffled link sets. The data is obtained from simulations using the actual link set of a random DNA configuration and then randomly reshuffling the positions of these links (while conserving the length of each link). The protein dynamics is simulated on this reshuffled but temporally fixed link set. Finally the average dynamics of the width $\Lambda(t)$ is obtained by averaging over many initial DNA configurations (each randomly drawn). The dynamics of $\Lambda(t)$ is super-diffusive, showing that the temporal correlations are not sufficient to produce the quasi-diffusive behavior (see main text).

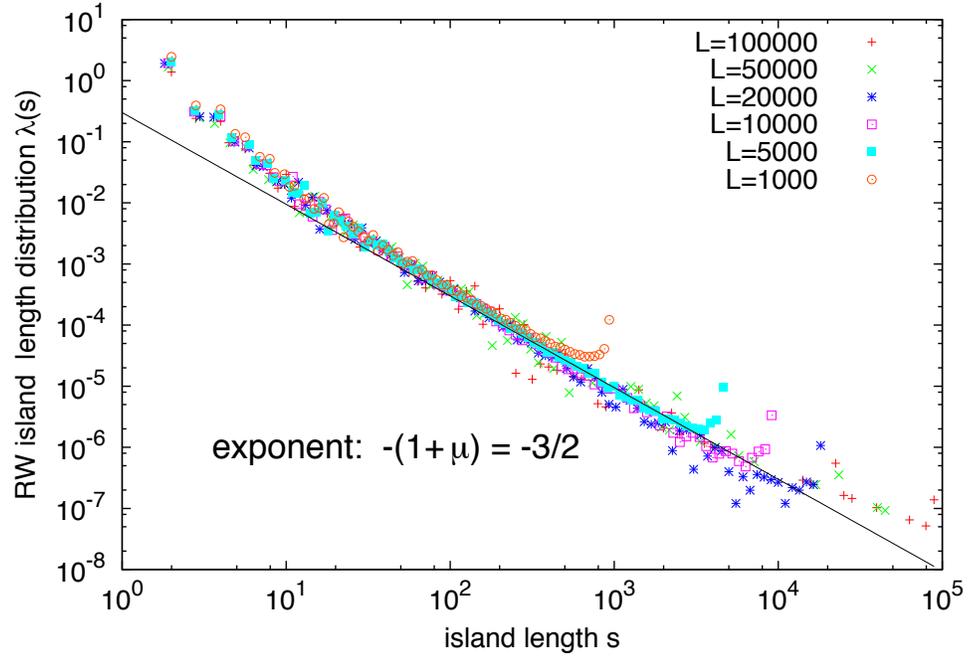

FIG. S5: Distribution of islands sizes. The distribution was obtained by generating random DNA conformations of length $L$ and picking a single island from each link diagram at random. The distribution displays the same power law decay as the distribution of link lenghts.

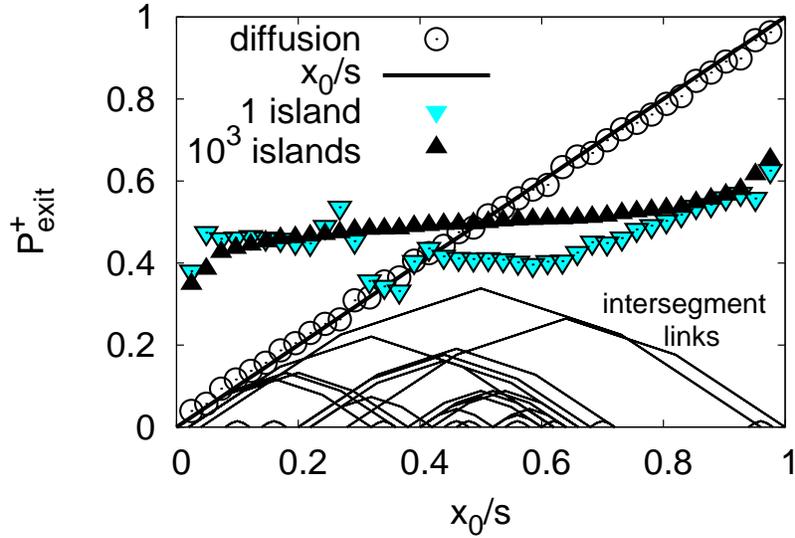

FIG. S6: Exit probability from an island to the right as a function of the starting position $x_0$ (normalized here by the island size $s$). The exit probability is shown both for a single randomly chosen island (blue downward triangles) with the link configuration indicated by the link diagram in the bottom, as well as averaged over an ensemble of 1000 islands of the same size (black triangles). For comparison, the case of pure diffusion (no links), where the exit probability depends linearly on the starting position, is also shown (the solid line shows the analytical dependence while the circles indicate simulation data, which was obtained as a control using the same simulation code as for the islands). It is evident that the probability of exiting an island on a given side depends only weakly on the initial position, at least in the core of the island. This justifies our coarse grained hopping model in island space.

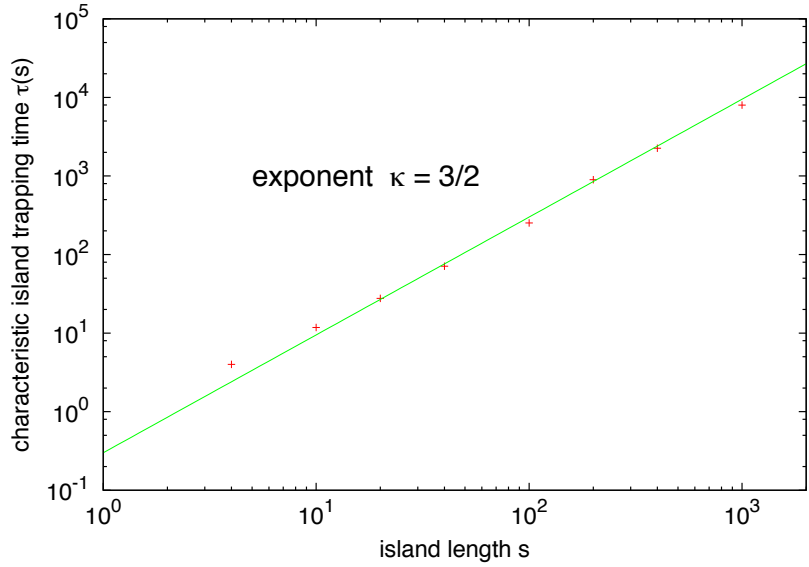

FIG. S7: Average trapping time of a protein within an island as a function of island size $s$. Each data point is obtained as an average over many simulations where a protein is initialized in the center of an island of size $s$ within a randomly drawn DNA configuration, and the time until it exits from the island is recorded. This island-size dependent characteristic trapping time scales as $\tau(s) \sim s^{3/2}$.


\end{document}